\documentclass{article}
\usepackage{html}

\title{A system for reflection in C++}
\author{Duraid Madina and Russell K. Standish\\
High Performance Computing Support Unit\\
University of New South Wales\\Sydney, 2052\\Australia\\
\{duraid,R.Standish\}@unsw.edu.au\\http://parallel.hpc.unsw.edu.au/rks}

\newcommand{\EcoLab}{{\sffamily\slshape
    \mbox{\raisebox{.5ex}{Eco}\hspace{-.4em}{\makebox[.5em]{L}ab}}}}

\begin{document}
\maketitle

\begin{abstract}
Object-oriented programming languages such as Java and Objective C have
become popular for implementing agent-based and other object-based
simulations since objects in those languages can {\em reflect} (i.e. make
runtime queries of an object's structure). This allows, for example, a
fairly trivial {\em serialisation} routine (conversion of an object into a
binary representation that can be stored or passed over a network) to be
written. However C++ does not offer this ability, as type information is
thrown away at compile time. Yet C++ is often a preferred development
environment, whether for performance reasons or for its expressive features
such as operator overloading.

In this paper, we present the {\em Classdesc} system which brings many
of the benefits of object reflection to C++.
\end{abstract}

\section{Classdesc}\label{obj-desc}

Object {\em reflection} allows straightforward implementation of
serialisation (i.e. the creation of binary data representing objects that can
be stored and later reconstructed), binding of scripting
languages or GUI objects to `worker' objects and remote method
invocation.  Serialisation, for example, requires knowledge of the
detailed structure of the object. The member objects may be able to
serialised (eg a dynamic array structure), but be implemented in terms
of a pointer to a heap object. Also, one may be interested in
serialising the object in a machine independent way, which requires
knowledge of whether a particular bitfield is an integer or floating
point variable. 

Languages such as Objective C give objects reflection by creating
class objects and implicitly including an {\em isa} pointer in objects
of that class pointing to the class object.  Java does much the same thing,
providing all objects with the native (i.e. non-Java) method {\tt getClass()}
which returns the object's class at runtime, as maintained by the virtual
machine.

When using C++, on the other hand, at compile time most of the information
about what exactly objects are is discarded. Standard C++ does provide a run-time
type enquiry mechanism, however this is only required to return a
unique signature for each type used in the program. Not only is
this signature be compiler dependent, it could be implemented by the
compiler enumerating all types used in a particular compilation,
and so the signature would differ from program to program!

The solution to this problem lies (as it must) outside the C++ language per
se, in the form of a separate program which parses an input program and
emits function declarations that know about the structure of the objects
inside. These are generically termed {\em object descriptors}. The object
descriptor generator only needs to handle class, struct and union
definitions. Anonymous structs used in typedefs are parsed as well.  What is
emitted in the object descriptor is a sequence of function calls for each
base class and member, just as the compiler generated constructors and
destructors are formed. Function overloading ensures that the correct
sequence of actions is generated at compile time.

For instance, assume that your program had the following class
definition:
\begin{verbatim}
class test1: base_t
{
  int x,y;
public:
  double z[100];
};
\end{verbatim}
and you wished to generate a serialisation operator called {\tt pack}.
Then this program will emit the following function declaration for
{\tt test1}:
\begin{verbatim}
#include "pack_base.h"
void pack(pack_t *p, string nm, test1& v)
{
   pack(p,nm,(base_t)v);
   pack(p,nm+".x",v.x);
   pack(p,nm+".y",v.y);
   pack(p,nm+".z",v.z,100);
}
\end{verbatim}
  
Thus, calling {\tt pack(p,"",var)} where {\tt var} is of type {\tt test1},
will recursively descend the compound structure of the class type,
until it reaches primitive data types which can be handled by the
following generic template:
\begin{verbatim}
template <class T>
void pack(pack_t *p,string desc, T& arg)
{p->append((char*)&arg,sizeof(arg));}
\end{verbatim}
given a utility routine {\tt pack\_t::append} that adds a chunk of data to a
repository of type {\tt pack\_t}.

This can even be given an easier interface by defining the member
template:
\begin{verbatim}
template <class T>
pack_t& pack_t::operator<<(T& x) 
{::pack(this,"",x);}
\end{verbatim}
so constructions like \verb+buf << foo << bla;+ will pack the objects
{\tt foo} and {\tt bla} into the object {\tt buf}. 

At the time of writing, this system has been implemented as a part of
the \htmladdnormallinkfoot{\EcoLab{} simulation
system}{http://parallel.hpc.unsw.edu.au/rks/ecolab}. The pack and
unpack operations work more or less as described. The type
\verb+xdr_pack+, derived from \verb+pack_t+ uses the standard unix XDR
library to pack the buffer in a machine independent way. This allows
checkpoint files to be transported between machines of different
architectures, or to run the simulation in a client-server mode, with
the client downloading a copy of the simulation whilst the simulation
is in progress.

Also implemented is a \verb+TCL_obj+ class descriptor. When
\verb+TCL_obj+ is called on an object, TCL\footnote{TCL is a well
known scripting language --- see http://dev.scriptics.com} commands
are created in the TCL interpreter that allow the C++ object's members
to be queried or set directly from the script. Moreover, provided
member functions take the arguments \verb+(int argc,char *argv[])+ or
no arguments at all, member functions can be called directly from the
TCL script.

The Classdesc software can be used in other projects by ``installing''
\EcoLab{} into a well known location, and using the classdesc command to
generate the descriptors. It is planned to separate out a
``slimmed-down'' Classdesc product that leaves out most of the \EcoLab{}
support libraries.

\section{Subtleties}

\subsection{Pointers}

Pointers create difficulties for Classdesc, since pointers may point
to a single object, an array of objects, functions, members or even
nothing at all. When array sizes are known at compile time, Classdesc
issues an object descriptor that loops over the elements, however
arrays allocated dynamically on the heap through the use of {\tt new}
cannot be handled, even in principle. It turns out that one can
distinguish between member pointers and normal pointers quite easily
through overloading of object descriptors, however it proves
impossible to distinguish between pointers to functions and pointers
to objects at the overloading stage. It is too difficult for Classdesc
to distinguish between function and object pointers, since the type
definitions may be hidden behind typedefs in other files. All that can
be done is to ignore pointer members and issue a runtime warning that
an attempt was made to serialise a generic pointer.

This strategy misses an important use of pointers in representing data
structures such as graph and trees. In this case, we may assume that
pointers point to a valid object, or are {\tt NULL}. However, we must
ask the programmer to flag the data types that have this
property. Currently, this is implemented in the {\tt pack} descriptor
by a macro \verb+Single_Obj_Ptr+, which takes the typename as an
argument. This has the disadvantage of polluting global namespace, and
requiring different macros for different descriptors. An alternative
approach (not yet implemented) is to ask the programmer to derive the
datatype from a particular given type (let's say
\verb+single_obj_ptr+). Then the generic pointer descriptor can
attempt to do a \verb+dynamic_cast+ --- if the cast succeeds, the
pointer points to a valid object, or is NULL\footnote{at least that is
what the programmer is promising}, and if the cast fails, then the
pointer may be anything, and should be ignored.

\subsection{Dynamic Types}

Dynamic data types pose a special problem for Classdesc, since the
actual storage is usually separate from the variable representing
it. The standard method of handling dynamic types inherited from C is
via pointers to memory allocated by {\tt new} or {\tt malloc()}. As
mentioned in the previous section, Classdesc cannot handle this method
of implementing dynamic data. In a more general sense, this method
should be discouraged, since it is very easy to create dangling
pointers and memory leaks. 

In C++, a better method is to encapsulate a dynamic type in a
class. This has the following advantages:
\begin{enumerate}
\item Calls to {\tt new} and {\tt delete} can be properly balanced, by
ensuring the {\tt delete} appears in the destructor.
\item Usual pointer syntax can be preserved by defining
\verb+operator[](int)+, \verb+operator*()+,  \verb+operator+(int)+,
\verb"operator++(int)", with the possibility of runtime range checking
built in.
\item Specialised descriptors can be supplied for this class, which
will integrate neatly with those produced by Classdesc. One may use the pragma
{\tt \#pragma omit} to indicate that a particular type
already has a descriptor defined.
\end{enumerate}

\subsection{Member privacy}

Descriptors need access to all members of an object, including private
and protected ones. Since in C++ class namespaces are closed by design
(no new members can be added, except by inheritance), descriptors need
to be placed a global or an open namespace. This means that friend
declarations need to be added to all class definitions with private or
protected areas. This is handled by an auxiliary program {\tt
insert-friend} that inserts the macro call {\tt
CLASSDESC\_ACCESS(}{\em class name}{\tt )} into class definitions, and
\linebreak[4] {\tt
CLASSDESC\_ACCESS\_TEMPLATE(}{\em class name}{\tt )} into template
class definitions. The intention is for the programmer to define macros
similar to the following:
\begin{verbatim}
#define CLASSDESC_ACCESS(type)\
friend void pack(pack_t *,eco_string,type&);\
friend void unpack(unpack_t *,eco_string,type&);

#define CLASSDESC_ACCESS_TEMPLATE(type)\
friend void pack<>(pack_t *,eco_string,type&);\
friend void unpack<>(unpack_t *,eco_string,type&);
\end{verbatim}

The script {\tt fix-includes} parses all standard header libraries,
and if private or protected members exist within that header, places a
modified version with the CLASSDESC macros into a separate include
directory. This works reasonably well in practice, provided the
authors of the header files have not done anything too tricky with
macro processing.

\section{Current Status}\label{conclusion}

Classdesc is currently a part of the \EcoLab{} simulation system.  \EcoLab{} is an
open source project, and is available from the
\htmladdnormallinkfoot{\EcoLab{}
website}{http://parallel.hpc.unsw.edu.au/rks/ecolab}.  Classdesc is
contained within a subdirectory of \EcoLab, so it is planned that at
distribution time, a separate Classdesc distribution be created for
each \EcoLab{} distribution.

There is an email discussion list, which you can subscribe to by
sending the message ``{\tt subscribe ecolab-list}'' to
majordomo@explode.unsw.edu.au. You can also register as a developer by
emailing R.Standish@unsw.edu.au. This will allow you to access the
code as it is being developed, and submit your own code
changes. \EcoLab{} is managed by Peter Miller's \htmladdnormallinkfoot{Aegis}
{http://www.canb.auug.org.au/\~{}millerp/aegis/aegis.html} code
management system, which allows multiple developers to work
independently on the code.

\end{document}